\documentclass{article}
\usepackage{spconf,amsmath,graphicx}

\usepackage{enumitem}
\setlist{nosep, leftmargin=14pt}

\usepackage{mwe} 
\usepackage{amssymb}

\title{Cross-Modal Fine-Tuning of 3D Convolutional Foundation Models for ADHD Classification with Low-Rank Adaptation}

\name{%
  \parbox{\linewidth}{\centering
  Jyun-Ping Kao$^{1,2}$, Shinyeong Rho$^{1,3}$, Shahar Lazarev$^{4}$, Hyun-Hae Cho$^{1,5}$, Fangxu Xing$^{1}$,\\
  {\itshape  Taehoon Shin$^{5}$, C.-C. Jay Kuo$^{6}$, Jonghye Woo$^{1}$}
  }
}
\address{%
  \makebox[\textwidth][c]{$^{1}$Massachusetts General Brigham and Harvard Medical School, Boston, MA, USA}\\
  \makebox[\textwidth][c]{$^{2}$National Taiwan University, Taipei, Taiwan}\\
  \makebox[\textwidth][c]{$^{3}$Washington University in St. Louis, MO, USA}\\
  \makebox[\textwidth][c]{$^{4}$Tel Aviv University, Tel Aviv, Israel}\\
  \makebox[\textwidth][c]{$^{5}$Ewha Womans University, Seoul, Korea}\\
  \makebox[\textwidth][c]{$^{6}$University of Southern California, Los Angeles, CA, USA}\\
}

\begin{document}
%
\maketitle
\begin{abstract}
Early diagnosis of attention-deficit/hyperactivity disorder (ADHD) in children plays a crucial role in improving outcomes in education and mental health. Diagnosing ADHD using neuroimaging data, however, remains challenging due to heterogeneous presentations and overlapping symptoms with other conditions. To address this, we propose a novel parameter-efficient transfer learning approach that adapts a large-scale 3D convolutional foundation model, pre-trained on CT images, to an MRI-based ADHD classification task. Our method introduces Low-Rank Adaptation (LoRA) in 3D by factorizing 3D convolutional kernels into 2D low-rank updates, dramatically reducing trainable parameters while achieving superior performance. In a five-fold cross-validated evaluation on a public diffusion MRI database, our 3D LoRA fine-tuning strategy achieved state-of-the-art results, with one model variant reaching 71.9\% accuracy and another attaining an AUC of 0.716. Both variants use only 1.64 million trainable parameters (over 113× fewer than a fully fine-tuned foundation model). Our results represent one of the first successful cross-modal (CT-to-MRI) adaptations of a foundation model in neuroimaging, establishing a new benchmark for ADHD classification while greatly improving efficiency.

\end{abstract}
\section{Introduction}
\label{sec:intro}

Attention-deficit/hyperactivity disorder (ADHD) is one of the most common neurodevelopmental disorders, affecting approximately 5–10\% of children worldwide \cite{shaw2009new}. Early diagnosis is crucial for improving educational, social, and mental health outcomes \cite{hamed2015diagnosis}. However, current ADHD diagnosis relies heavily on subjective behavioral assessments that can be inconsistent and prone to bias, leading to variable clinical decisions. Magnetic resonance imaging (MRI) offers a noninvasive window into brain structure and function, but existing MRI-based ADHD classifiers have achieved only modest performance (area under the curve (AUC) $\approx 0.7$ \cite{zhou2021multimodal, lin2023population, chiang2024machine}) by incorporating structural MRI, resting-state fMRI, and diffusion MRI features, emphasizing their limited discriminative power and the need for improved methods. 

Recently, large-scale foundation models (FM) pre-trained on medical images have shown promise in learning generalizable representations \cite{pai2024foundation}. For example, a 3D ResNet-50 FM trained on 11,467 computed tomography (CT) lesion volumes \cite{pai2024foundation} achieved robust performance in oncology imaging. We hypothesize that such a high-capacity CT-based model can be repurposed to detect subtle neuroanatomical patterns in a different modality and domain despite the modality and clinical context differences. 

A key challenge lies in the model’s scale: fine-tuning hundreds of millions of parameters on a small ADHD MRI database is computationally impractical and prone to overfitting. Parameter-efficient fine-tuning (PEFT) techniques address this by updating only a tiny fraction of parameters. One leading PEFT method is Low-Rank Adaptation (LoRA) \cite{hu2022lora}, which injects small trainable low-rank matrices into a frozen model to capture task-specific updates, drastically reducing trainable parameters. While LoRA has been successfully applied to linear and attention layers in transformer-based architectures for medical segmentation and detection tasks \cite{zhang2023customized, kao2025lora}. However, applying LoRA to the 3D convolutional kernels of a volumetric convolutional neural network (CNN) is non-trivial, as 3D filters are high-dimensional tensors. To our knowledge, this gap has not been previously addressed. In this work, we address this gap by developing a 3D LoRA-based parameter-efficient fine-tuning approach for cross-modal adaptation of a FM, enabling CT-to-MRI knowledge transfer for ADHD classification.

\section{Method}
\label{sec:method}

\subsection{Foundation Model}

\begin{figure}[htb]
\begin{minipage}[b]{1.0\linewidth}
  \centering
  \centerline{\includegraphics[width=8.5cm]{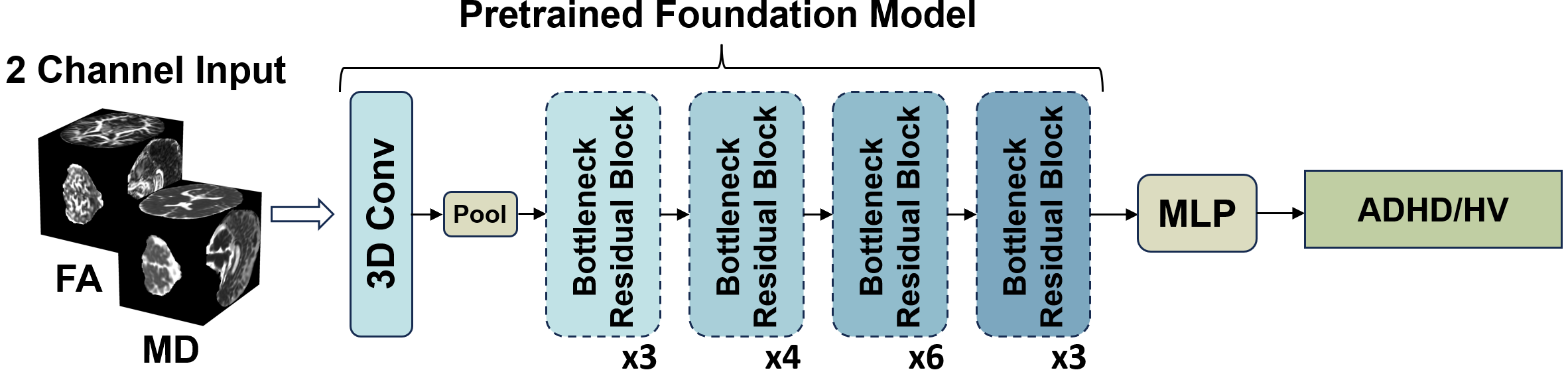}}
  \centerline{}\medskip
\end{minipage}
\caption{
Overview of the proposed ADHD classification pipeline. Input: A 3D tensor with two channels (FA and MD). Backbone: The input is processed by a pre-trained 3D FM \cite{pai2024foundation}. Multilayer Perceptron (MLP): Performs the final binary classification for ADHD and Healthy Volunteer (HV).}
\end{figure}

We leveraged a 3D ResNet-50 based FM for cancer imaging biomarkers (FMCIB) \cite{pai2024foundation} pre-trained on 11,467 CT lesion volumes as our fixed encoder backbone. To efficiently adapt this large model to the small ADHD MRI dataset, we inserted LoRA adapters into every 3D convolutional layer of the model.

\subsection{Low-Rank Adaptation for 3D Convolutions}
\begin{figure}[htb]
\begin{minipage}[b]{1.0\linewidth}
  \centering
  \centerline{\includegraphics[width=8.5cm]{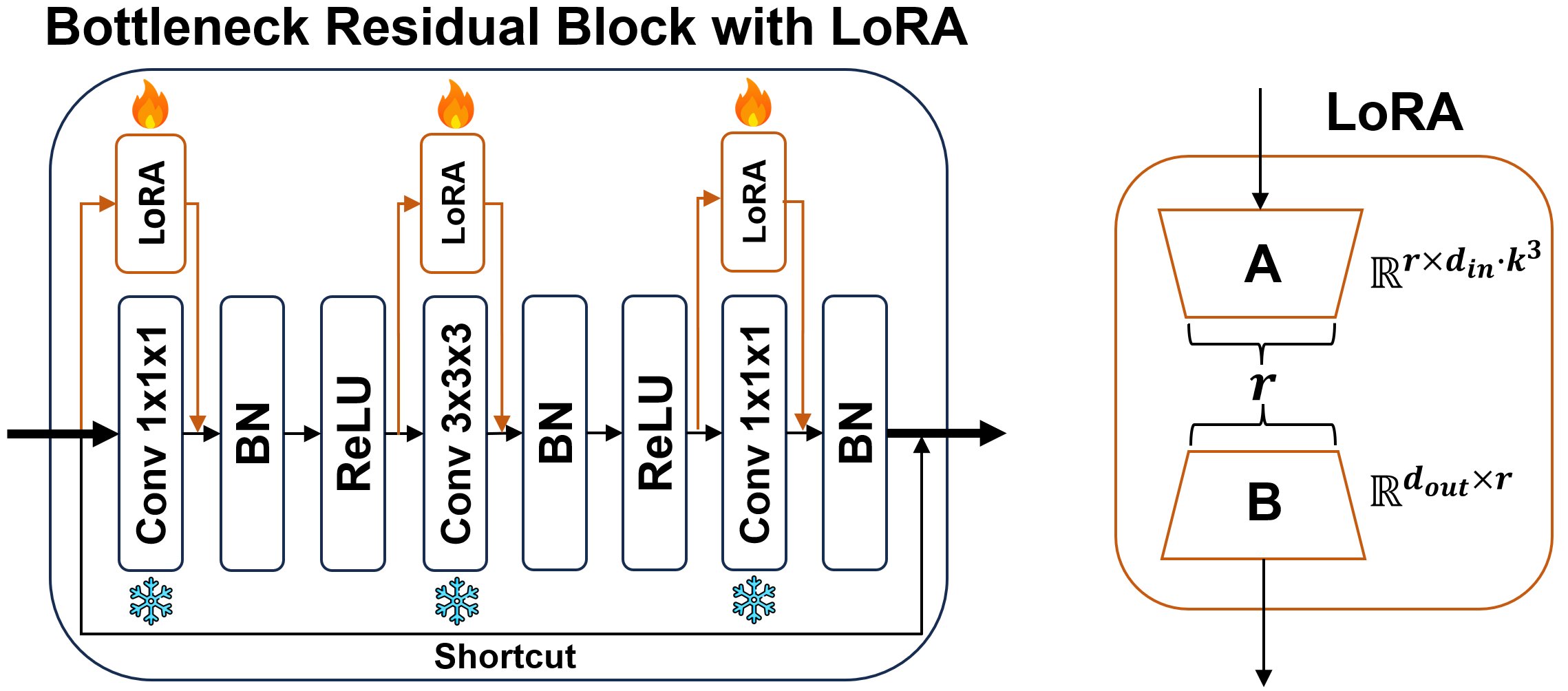}}
  \centerline{}\medskip
\end{minipage}
\caption{Illustration of the proposed LoRA applied within a bottleneck residual block. Left: The standard residual block architecture. Our trainable LoRA modules are injected in parallel to each 3D convolutional layer. Right: A detailed view of a LoRA module. }
\end{figure}
Each LoRA module comprises a pair of trainable matrices of rank $r=4$ that introduce an additive low-rank update to the convolutional filters. For a 3D convolutional with weight tensor $W\in\mathbb{R}^{d_{\text{out}}\times d_{\text{in}}\cdot k^3}$ (where $k$ is the spatial kernel's size), we parameterize the adapted weights as $W' = W + \Delta W=W+B A$, with $B\in\mathbb{R}^{d_{\text{out}}\times r}$ and $A\in\mathbb{R}^{r\times d_\text{in}\cdot k^3}$. In implementation, each 3D convolutional kernel is effectively reshaped so that the low-rank matrices act across two of its spatial dimensions, yielding a 2D adaptation per filter. We insert such a LoRA adapter in every 3D convolutional layer of the encoder. Thus, each layer’s output is the sum of the frozen convolution and the low-rank update. All LoRA parameters (the entries of $A$ and $B$) are trainable, while the original convolutional weights $W$ are kept fixed.

At last, a lightweight classifier maps the features to the ADHD prediction. The head is a two-layer MLP: the first layer has 128 hidden units with a Gaussian Error Linear Unit (GELU) activation and dropout (rate 0.5), and the second layer outputs the final logit for binary classification. Only the MLP head’s weights and the LoRA parameters are optimized during training. Thus, the final model consists of the frozen CT pre-trained 3D backbone, trainable LoRA modules in each convolutional layer, and a trainable MLP classifier.
\section{Experiments}
\label{sec:exp}
\begin{table*}[!t]
  \caption{Five-fold cross-validated mean performance for the MRI-based classification task, comparing Machine Learning (ML), Deep Learning (DL), and foundation-model (FM) approaches. Best results are bolded; second-best are underlined.}
  \label{tab:main-result-wide}
  \centering
  \small                             
  \setlength{\tabcolsep}{4pt}         
  \renewcommand{\arraystretch}{1.1}   
  \begin{tabular}{lcccccc}
    \hline
    Category & Model & Pre-trained Modality & Accuracy & AUC & Trainable Params (M) & FLOPs (T)\\
    \hline
    
    ML & Logistic Regression & From Scratch & 0.467 & 0.490 & -- & -- \\
    ML & Support Vector Machine & From Scratch & 0.437 & 0.481 & -- & -- \\
    ML & Decision Tree & From Scratch & 0.593 & 0.585 & -- & -- \\
    ML & Random Forest & From Scratch & 0.526 & 0.459 & -- & -- \\
    ML & XGBoost & From Scratch & 0.511 & 0.505 & -- & -- \\
    DL & 3D U-Net \cite{cciccek20163d} & From Scratch & 0.574 & 0.673 & 19.07 (11.63 $\times$) & 0.14 \\
    DL & 3D U-Net (Models Genesis) \cite{zhou2021models} & CT & \underline{0.600} & 0.557 & 19.34 (11.79 $\times$) & 1.96 \\
    DL & 3D ResNet-50 \cite{hara2018can} & From Scratch & 0.593 & 0.710 & 46.18 (28.16 $\times$) & 0.09 \\
    DL & 3D ResNet-50 (MedicalNet) \cite{chen2019med3d}  & CT + MRI & \underline{0.600} & 0.579 & 46.18 (28.16 $\times$) & 0.37 \\
    FM & FMCIB \cite{pai2024foundation} + MLP & CT & 0.556 & \underline{0.711} & 185.57 (113.15 $\times$) & 0.41 \\
    
    \hline
    \textbf{FM}& \textbf{ FMCIB \cite{pai2024foundation} + MLP + LoRA (Ours)} & CT & \textbf{0.719} & 0.630  & \textbf{1.64} & 0.41 \\
    \textbf{FM}& \textbf{ FMCIB \cite{pai2024foundation} + MLP + LoRA (Ours)} & CT & 0.630 & \textbf{0.716}  & \textbf{1.64} & 0.41 \\

    \hline
  \end{tabular}
\end{table*}

\subsection{Dataset and Preprocessing}

    

This study utilized the open source ``Emotion and Development Branch Phenotyping and DTI" \cite{dataset} dataset, which comprises a transdiagnostic sample of youth participants aged 7 to 20 years. The original cohort was recruited with inclusion criteria for individuals with ADHD, Disruptive Mood Dysregulation Disorder (DMDD), and typically developing Healthy Volunteers (HV). Diffusion MRI data were preprocessed using QSIPrep's default reconstruction pipeline~\cite{cieslak2021qsiprep}, including susceptibility distortion correction, head motion and eddy current correction through FSL's eddy tool~\cite{Andersson2016}, and registration to native anatomical space~\cite{Avants2011}. Diffusion tensor imaging (DTI) model fitting was performed on the corrected data, yielding fractional anisotropy (FA) and mean diffusivity (MD) maps that quantify white matter microstructural integrity in subject-specific space. Lastly, the data is resized into 128×128×128 for training. In total, the dataset comprised 129 cases, including 76 labeled ADHD (58.9\%) and 53 labeled HV (41.1\%). 

\subsection{Experimental Settings and Evaluation}
We trained the model using five-fold cross-validation (80/20 split in each fold) to maximize usage of the limited data. Each fold's model was trained with a binary cross-entropy loss using the AdamW optimizer. The learning rates for the LoRA parameters and the classification head were set to $10^{-4}$ and $10^{-5}$, respectively, with a weight decay of $10^{-4}$ applied to both. Machine learning baselines were established using the Python scikit-learn 1.7.2 library. We evaluated binary ADHD classification using overall accuracy and the area under the receiver operating characteristic curve (AUC). Accuracy measures the proportion of correct predictions  $\mathrm{Accuracy}=\frac{TP + TN}{TP + TN + FP + FN}$, where $TP, TN, FP, FN$ are entries of the confusion matrix. We saved model checkpoints throughout training and later selected two representative fine-tuned models: one chosen for highest validation accuracy and another for highest validation AUC. 
For probabilistic outputs, the model produces a score $s_i\in[0,1]$ for the ADHD class from the softmax over logits, and the ROC curve is obtained by sweeping a threshold $\tau$ on $s_i$.~With $\mathrm{TPR}(\tau)=\frac{TP}{TP+FN}$ and $\mathrm{FPR}(\tau)=\frac{FP}{FP+TN}$, AUC is the integral of true positive rate with respect to false positive rate.

\section{Results}
\label{sec:exp}


Our proposed framework fine-tunes a pre-trained FM via our novel LoRA adaptation strategy, achieving substantially better performance than all conventional baselines in ADHD classification. Table \ref{tab:main-result-wide} details the five-fold cross-validated mean performance of Machine Learning, Deep Learning (training from scratch or with medical-image pretraining), and our LoRA-based FM fine-tuning approach. Classical ML classifiers yielded near-chance results with AUC values around 0.50, underscoring the difficulty of this task when using simple models. DL models trained from scratch performed better in terms of AUC, achieving AUCs of 0.67-0.71, while transfer learning with medical imaging pre-trained weights yielded AUCs of only 0.56–0.58. This outcome suggests that without a specialized adaptation strategy, much of the pre-trained knowledge cannot be effectively transferred across imaging modalities. We report two model variants: one optimized for maximal ACC and another for maximal AUC. The ACC-optimized variant achieved an accuracy of 0.719, significantly outperforming the best prior deep model's accuracy of 0.600. The AUC-optimized variant achieved an AUC of 0.716, exceeding the strongest baseline of 0.711. Importantly, both models achieve superior performance while fine-tuning only 1.64 million parameters, which is 113× fewer than fully fine-tuning the entire FM. This remarkable efficiency highlights the effectiveness of our 3D LoRA strategy in preserving and transferring the FM’s knowledge.

\section{Discussion and Conclusion}
\label{sec:dc}

In this work, we introduced a novel application of LoRA in 3D convolutional kernels, demonstrating that a highly PEFT strategy can successfully transfer knowledge from a CT-based FM to an MRI-based ADHD classification task. This cross-modal adaptation achieved new state-of-the-art results for ADHD diagnosis using only two diffusion MRI-derived feature maps, including FA and MD, as input. 

The success of our approach carries several important implications. First, we validated the feasibility of ambitious cross-modal (CT-to-MRI) and cross-domain (oncology-to-psychiatry) transfer learning using FM. Our results strongly support the notion that FMs pre-trained on diverse medical images learn fundamental, modality-agnostic representations that can generalize across imaging modalities and clinical domains. In this case, features learned from CT scans in an oncology context proved robust enough to detect the subtle microstructural brain alterations characteristic of ADHD in MRI. This represents one of the first demonstrations that a medical imaging FM can be fine-tuned across modalities to solve a neuropsychiatric classification problem, demonstrating the broad potential of cross-domain adaptation. 

 Second, the divergent behavior of our ACC- and AUC-optimized models highlight a crucial consideration in clinical model development. The ACC-optimized model, by concentrating prediction probabilities around the decision boundary, is well suited for direct diagnostic use (maximizing immediate classification accuracy at the 0.5 threshold). 
 In contrast, though poorly calibrated for thresholds, the AUC-optimized model excels at risk ranking, making it a valuable risk-stratification tool for prioritizing individuals likely to have ADHD for review. Despite the promising results, we acknowledge that our experiments were conducted on a single publicly available dataset, and validating the approach on larger, multi-site cohorts is needed to ensure generalizability across diverse populations and imaging conditions.


%
%
%


\section{COMPLIANCE WITH ETHICAL STANDARDS}
This research study was conducted retrospectively using human subject data made available in open access by \cite{dataset}.

\section{Acknowledgments}
The authors thank Dr. Xiaofeng Liu for insightful discussions. This work is partly supported by the National Research Foundation of Korea funded by the Ministry of Science and ICT (RS-2024-00338438).

\bibliographystyle{IEEEbib}
\bibliography{refs}
\end{document}